\documentstyle[aps,prl,epsfig,multicol]{revtex} 
\newcommand{\Beginfigure}{\vskip 0pt\noindent}
\newcommand{\Endfigure}{\vskip 0pt\noindent}

\newcommand{\Caption}[1]{\par\noindent{\footnotesize #1} \par} 

\newcommand{\Beginrule}{\vskip 3pt\noindent\hbox{%
\vbox{\hbox to 9cm{\hfill}}\vbox{\hrule width 9cm}} \vskip 3pt}
\newcommand{\infig}[2]{\begin{center}{\epsfig{file=#2,width=#1}}\end{center}} 
\begin{document}

\title{Quantum Kinetic Theory of Condensate Growth---Comparison of 
Experiment and Theory}
\author{C.W.~Gardiner$^{1,4}$ M.D. Lee$^{1,4}$  R.J. Ballagh$ ^{2}$ 
M.J. Davis$ ^{2}$ 
and P. Zoller$ ^{3,4}$.}
\address{$^1$ School of Chemical and Physical Sciences, 
Victoria University, Wellington, New Zealand}
\address{$^2$ Physics Department, University of Otago, Dunedin, New 
Zealand}
\address{$^3$ Institut f\"ur theoretische Physik, Universit\"at 
Innsbruck, A6020 Innsbruck, Austria.} 
\address{$^4$ Institute for Theoretical Physics, University of 
California, Santa Barbara, CA 93106-4030, U.S.A.} \maketitle

\begin{abstract}
 In a major extension of our previous model \cite{BosGro} of 
 condensate growth, we take account of the evolution of the 
 occupations of lower trap levels, and 
 of the full Bose-Einstein formula for the occupations of higher trap 
 levels.  We find good agreement with experiment, especially at higher 
 temperatures.  We also confirm the picture of the ``kinetic'' region 
 of evolution, introduced by Kagan {\em et al.}, 
 for the time up to the initiation of the condensate.  The behavior 
 after initiation essentially follows our original growth equation, 
 but with a substantially increased rate coefficient $W^{+}$.
\end{abstract}

\pacs{PACS Nos. }
\begin{multicols}{2}


 Although the first Bose condensed atomic vapor was produced in a 
 magnetic trap only in 1995 \cite{JILA,MIT,RICE}, the kinetics of 
 condensate formation has long been a subject of theoretical study 
 \cite{Older,Kagan92}.  There is now intense theoretical work on 
 Bose-Einstein condensation, which is excellently summarised in 
 \cite{Stringari98}.  Most theoretical studies of condensate growth 
 either have not treated trapping, or have considered only traps which 
 are so broad that the behavior of the vapor is not essentially 
 different from the untrapped situation.  Furthermore, they have given 
 only {\em qualitative} estimates of condensate growth.  Our previous 
 paper \cite{BosGro} introduced a simplified formula for the growth of 
 a Bose-Einstein condensate, in which growth resulted from stimulated 
 collisions of atoms in a thermal reservoir, where one of the atoms 
 enters the lowest trap eigenstate, whose occupation thus grows to 
 form a condensate.  We thus included the trap eigenfunctions as an 
 essential part of our description, and gave the first {\em 
 quantitative} formula for the growth of a condensate.  The growth 
 rate was of the order of magnitude of that estimated from experiments 
 current at that time.

This direct stimulated effect must be very important once a 
significant amount of condensate has formed, but in the initial stages 
there will also be a significant number of transitions to other low 
lying trap levels whose populations will then also grow.  As well as 
this, there will be interactions between the condensate, the atoms in 
these low lying levels, and the atomic vapor from which the condensate 
forms.  This paper will extend the description of the condensate 
growth to include these factors, and will compare the results with 
experimental data on condensate growth obtained at 
MIT~\cite{MITgrowth}, with which good agreement is found.  

As in our previous work, we divide the states in the potential into 
the {\em condensate band}, $ R_{\rm C}$, which consists of the energy levels 
significantly affected by the presence of a condensate in the ground 
state, and the {\em non-condensate band}, $ R_{\rm NC}$, which contains all the 
remaining energy levels above the condensate band.  The division 
between the two bands is taken to be at the value, $ e_{\rm max}$.  
The situation is illustrated in Fig. 1.
The picture we shall use assumes that $ R_{\rm NC}$ consists of a large  ``bath''
of atomic vapor, in thermal equilibrium, whose distribution function is given
by a time-independent equilibrium  Bose-Einstein distribution
$\{\exp[(E-\mu)/k_BT] -1\}^{-1}$ with positive chemical potential $\mu$. The
value  of $ e_{\rm max}$ will be assumed to be small enough for the majority 
of atoms to have energies higher than $ e_{\rm max}$, so that this part  of
the bath can be treated as being undepleted by the  process of condensate
growth.




\Beginfigure
\infig{8.5cm}{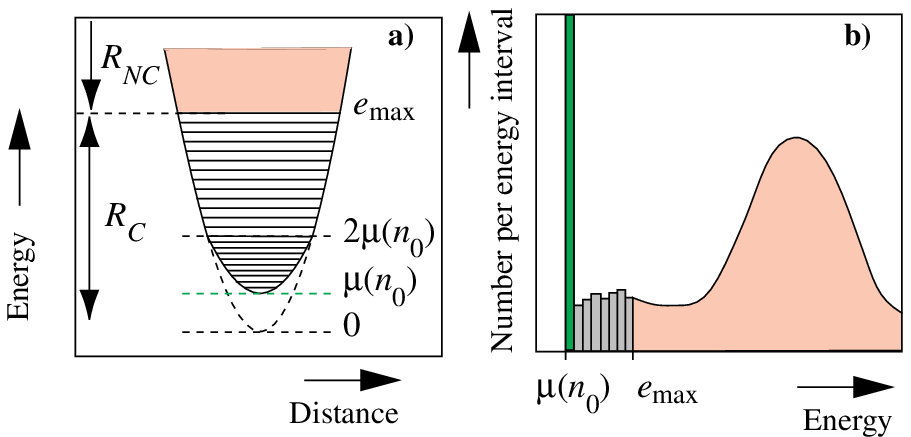} \Caption{Fig.~1: a) Representation of the 
change in the energy spectrum due to the growth of the condensate.  
b) Occupations of the levels considered - leftmost is the condensate 
level, followed by several discrete energy bands, with the 
constantly occupied $R_{\rm NC}$ at higher energies.  }
\Endfigure

In the following we will use the notation of $m$ as the mass of an 
atom with {\em s}-wave scattering length $a$, and $n_0$ for the number 
of condensate atoms.

The levels in $R_{\rm C}$ must have time dependent energies due to the effects of the
interaction with the  growing condensate. The energy of the ground state is
$\mu(n_0)$, the chemical potential, for which we use a modified Thomas-Fermi
approach $\mu(n_0) = \alpha (n_0 +\nu)^{2/5}$ where $\alpha =
(15a\omega_x\omega_y\omega_zm^{1/2}\hbar^2/4\sqrt{2})^{2/5}$ and $\nu$ is a
constant chosen so that $\mu(0) = \hbar(\omega_x+\omega_y+\omega_z)/2$.   As
$\mu(n_0)$ rises with an increase in $n_0$, the energies of higher energy
levels must also rise.  The exact nature of this rise does not affect the
results greatly.  An approximate treatment arises by leaving the levels with
energies above $2\mu(n_0)$ unchanged (including all levels in $R_{\rm NC}$), and
compressing the spacing of the levels under $2\mu(n_0)$, so that the energies
are given by
\begin{eqnarray} \label{levelenergies}
e_m = e_m^0 + 
\theta \big(2\mu(n_0) - e_m^0\big)\big(2\mu(n_0)-e_m^0-\mu(0)\big)/2
\end{eqnarray}
where $e_m^0$ are the non-interacting harmonic potential energy levels, and
$\theta(x)$ is the step function.
Note that $e^{0}_{0}=\mu(0)$, and thus $e_{0}=\mu(n_{0})$.
The levels used in this model are
represented graphically in Fig.~1.  To simplify the equations we also group the levels in narrow  bands of mean
energy $e_k$, $ g_k$ levels per group, and $ n_k$ particles per group.  This
corresponds to the {\em ergodic}  assumption used in 
\cite{Holland KE,SVISTUNOV,Kagan92}

Equations of motion then follow from Quantum Kinetic theory~\cite{QKIII} and
full derivation will be given elsewhere, but the result is essentially 
equivalent to modifying the Quantum Boltzmann (QBE) equation as 
follows.

{\bf (i)} We
use the QBE in an approximated ergodic form, where 
$e_{{\min}} =\min(e_{m},e_{n},e_{p},e_{q})$~\cite{Holland KE}
\begin{eqnarray}
\frac{\partial f(e_n)}{\partial \tau} 
&= &
\sum_{e_m,e_p,e_q}
\delta(e_m+e_n - e_p - e_q) g(e_{\rm min}) \times\nonumber \\
&& \Big[f(e_p)f(e_q) (1+f(e_m)) (1+f(e_n)) \nonumber \\
&&  \quad - f(e_m)f(e_n) (1+f(e_p)) (1+f(e_q))\Big],
\label{eqn:QBE}
\end{eqnarray}
where
$n_k = g_k\,f(e_k)$ is the number of particles with energy $e_k$,
and $\tau = (8ma^2\omega^2/\pi\hbar)\times t$.

{\bf (ii)} We use the modified energy levels as given above, but otherwise do not
change the QBE.

{\bf (iii)} We sum out over all levels in $R_{\rm NC}$ which is assumed
time independent.

{\bf (iv)} We omit any scattering between atoms which
both have energies less than $ e_{\rm max}$, which is reasonable if the number
of atoms in the bath is almost 100\% of the total number of atoms.

\Beginfigure
\infig{8.5cm}{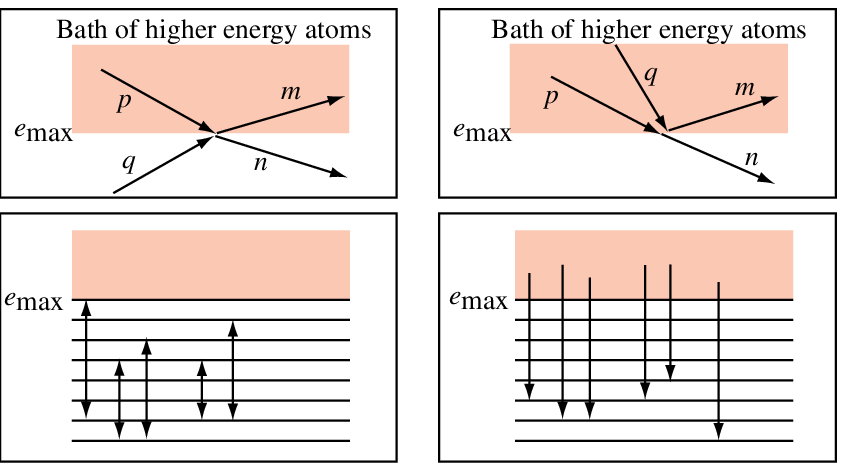}
\Caption{Fig.~2:  The transitions 
being considered: Left---scattering; Right---Growth.}
\Endfigure
As a result of this procedure there are two different kinds of dynamics, {\em
scattering} and {\em growth}, as illustrated in Fig.~2.  The evolution for the
population of the $m$th level in $R_{\rm C}$ is then
\begin{eqnarray}
{\partial n_m \over \partial \tau} &=&\dot{n}_m  = \dot{n}_m|_{\rm growth}+ 
\dot{n}_m|_{\rm scatt}
\label{growth1501}
\end{eqnarray}
where the individual terms are as follows.

\noindent
{\bf Scattering}: A collision between an atom initially in an energy 
level below $ e_{\rm max}$ and a bath atom transfers the first atom 
to another energy level below  $ e_{\rm max}$.  This is described by
\begin{eqnarray}
&&\left.{\dot n_m }\right |_{\rm scatt}
= e^{\mu/k_BT}\Gamma(T)\times
\nonumber\\ 
&&\,\,
\Bigg\{\sum_{k<m}{1\over g_m}\left[n_k(g_m+n_m)
{e^{-\hbar\omega_{mk}/k_BT} }
-n_m(g_k+n_k)\right]
\nonumber \\
 &&\,\,  +
\sum_{k>m}{1\over g_k }\left[n_k(g_m+n_m)
-n_m(g_k+n_k){e^{-\hbar\omega_{km}/k_BT}}\right]\Bigg\}.
\nonumber \\
\label{growth14}
\end{eqnarray}
where $\omega_{km}=e_{k}-e_{m}$.  The value of $ \Gamma(T) 
\equiv\sum_{e_m>e_{\rm max}}e^{-e_m/k_BT} $  depends on the 
energy spectrum.  We use the value for an {\em isotropic} 
3-dimensional harmonic oscillator with frequency $ \omega 
=(\omega_x\omega_y\omega_z)^{1/3}$.  Thus $ e_n = (n+3/2)\hbar\omega$, 
so that we find
\begin{eqnarray}\label{growth10a}
\Gamma(T) ={ e^{-e_{\rm max}/k_BT}\over1-e^{-\hbar\omega/k_BT}}.
\end{eqnarray}
However this value is not critical; similar results are obtained for 
any $\Gamma(T)$ greater than about 10\% of (\ref{growth10a}).


\noindent {\bf Growth}: A collision between a pair of atoms initially 
in the bath of atomic vapor results in {\em one} of the atoms having a 
final energy less than $ e_{\rm max}$.  This gives
\begin{eqnarray}
\dot{n}_m|_{\rm growth} &=& 2[(n_m+1) W^+_m(n_0) - n_m W^{-}_m(n_0)]
\end{eqnarray}
where
\begin{eqnarray}
&&{W}^+_m(n_0) = 
\frac{1}{2}\int_{e_{\rm max}}^{\infty}de_1
\int_{e_{\rm max}}^{\infty}de_2 \int_{e_{\rm max}}^{\infty}de_3
f(e_1) f(e_2)  \nonumber \\
&&\quad \times (1 + f(e_3))\delta(e_1 + e_2 - e_3 -e_m), \label{W+m0}\\
&&{W}^-_m(n_0) =  
\frac{1}{2}\int_{e_{\rm max}}^{\infty}de_1
\int_{e_{\rm max}}^{\infty}de_2 \int_{e_{\rm max}}^{\infty}de_3
(1 + f(e_1))   \nonumber \\
&&\quad \times (1+f(e_2))f(e_3)\delta(e_1 + e_2 - e_3 -e_m).
\end{eqnarray}
Because we assume the non-condensate band is in thermal equilibrium,
with temperature $T$ and chemical
potential $\mu$, we have
\begin{eqnarray}
{W}^-_m(n_0) &=&\exp\left({(e_m-\mu)\over k_B T}\right){W}^+_m(n_0)
\end{eqnarray}


The value of $ W_0^+$ (henceforth $W^+$) differs significantly from that used
previously \cite{BosGro} in that the integrals in (\ref{W+m0}) are only
performed over energy levels higher than $e_{\rm max}$ (previously over all
levels), and the Bose-Einstein distribution function is used over all
the range of integration (previously approximated by the Maxwell-Boltzmann
distribution).  With these changes we find that
\begin{eqnarray}\label{newWplus1}
W^+(n_0) &=& {1 \over 2}\left({k_BT\over\hbar\omega}\right)^2
\bigg\{[\log(1- z)]^2 
\nonumber \\
&& +  z^2\sum_{r=1}^{\infty}[ z\, z(n_0)]^r
[\Phi( z,1,r+1)]^2\bigg\},
\end{eqnarray}
where
\begin{eqnarray}\label{newWplus2}
 z &=& e^{\left(\mu-e_{\rm max}\over k_BT\right)} 
\ \ \ \ z(n_0) = e^{\left(\mu(n_0)-e_{\rm max}\over k_BT\right)} \label{newWplus3}.
\end{eqnarray}
The function $ \Phi$ is the {\em Lerch transcendent} \cite{Bateman}, defined by
$
\Phi(x,s,a) = \sum_{k=0}^\infty {x^k/ (a+k)^s}.
$
These changes result in a significant correction, increasing $W^+$ by about a
factor greater than three (dependent on $T$, $\mu$ and the trap parameters)
compared to that used previously in~\cite{BosGro}, and producing correspondingly
faster growth.


By making a further approximation, that $W_m^+(n_0) \approx W^+(n_0)$, the
calculations required are significantly simplified. We can do this 
because  the $W_m^{+}(n_0)$ are an average over all  the
levels contained in the $m$th group, and hence are  expected to be
of the same order of magnitude as $W^+(n_0)$.  As a validity check, it was
found that the effect on the  condensate growth rate was small when the
$W_m^{+}(n_0)$ were  altered by a factor in the range $0.5 - 2$.  We now
have, for the growth terms
\begin{eqnarray}\label{phys4}
\dot n_m|_{\rm growth} 
&=& 
2W^{+}(n_0)\left\{\left[1-e^{e_m-\mu \over k_{B}T}\right]n_m+g_m
\right\},
\nonumber \\
\\ \label{phys5}
\dot n_0|_{\rm growth}  &=& 
2W^+(n_0)\left\{\left[1-e^{\mu(n_0)-\mu\over k_{B}T}\right]n_0+1
\right\}.
\end{eqnarray}
The overall evolution of the system can now be found from the 
numerical solutions to the set of equations (\ref{growth1501}).  The 
parameters used are $\omega_x = \omega_y = 2\pi \times 82.3{\rm Hz}$ 
and $\omega_z = 2\pi \times 18 {\rm Hz}$, as in~\cite{MITgrowth}, and
$a=2.75{\rm nm} $,  $e_{\rm max} \approx 2.2\mu\big(n_0({\rm eq})\big)$, 
where $n_0({\rm eq})$ is the final equilibrium value of $n_{0}$, which 
is defined by
$\mu\big(n_{0}({\rm eq})\big) =\mu$, the chemical potential of the 
vapor.
The number of energy bands was set as large as possible, but it was 
required that there were at least four levels in the first group above 
the condensate, in order represent the fact that the levels are 
discrete.

The improvements to this model over that used in~\cite{BosGro}
speed up the condensate growth by up to an order of magnitude, depending on the
exact parameters, as anticipated in~\cite{BosGro}.  The major cause of this is the
correction to $W^+$ arising from the use of the correct Bose-Einstein distribution. 
The inclusion of the scattering terms does not change the overall rate of growth
substantially (which is dominated by the bosonic stimulated growth), but does speed
up the initial period of growth (dominated by spontaneous growth) and shortens the
time before the stimulated term becomes significant.  This gives a much
sharper initial growth curve as compared to the smoother S-bend curves
of~\cite{BosGro}.
In Fig.~3 we present an example of the results obtained for the growth of all the
bands in $R_{\rm C}$, in which a number of features may be seen.

{\bf (i)} The effect of the initial conditions used can be seen from the front
corner.  In this example the initial populations for all but the top ten bands
were set at zero, whilst the top ten bands had populations determined by the
same Bose-Einstein distribution function as for $R_{\rm NC}$.  The figure shows
that this initial condition is rapidly smoothed out by scattering 
processes.  Different initial conditions merely generate a small 
change in the {\em initiation time} defined below.

\Beginfigure
\infig{8.5cm}{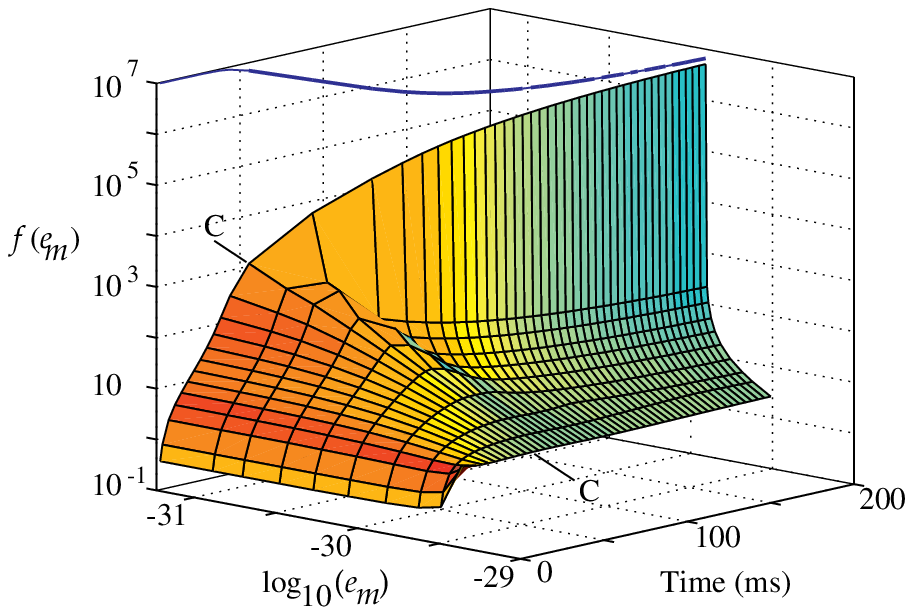} \Caption{Fig.~3: $f(e_{m})$ vs. 
energy $e_{m}$ as 
a function of time.  Note that the lines almost parallel to the time 
axis are not lines of constant energy, but rather lines of $f(e_{m})$ for 
constant level number whose energies change with condensate growth.
The solid curve in the plane at the top of the axes represents the 
curve $\log_{10}[e_{0}]\equiv\log_{10}[\mu(n_{0})]$ as a function of time.}
\Endfigure

{\bf (ii)} The {\em initiation time} (here $60{\rm ms}$) is defined by the critical
line C--C.  Up to this point the population of the condensate level is
relatively small.  The behavior up to the initiation time is similar to that 
found by Svistunov \cite{SVISTUNOV} and Kagan {\em et al.} 
\cite{Kagan92}. In particular the populations of the levels increase to 
approach a limiting dependence on energy of $f(E) \propto E^{-1.61}$ on 
the critical line C---C, which is in good agreement with their 
prediction of $E^{-5/3}$ \cite{SVISTUNOV HARMONIC} for the case of a 
harmonic trap.

{\bf (iii)} After the initiation time the condensate grows enormously.  
However, the occupations of the other trap levels actually decrease 
quite rapidly to their equilibrium values, and then remain nearly 
constant while the condensate continues to grow - by a factor of about 
10 in this case.  At the same time the energy spectrum of the trap 
levels in $R_{\rm C}$ changes according to (\ref{levelenergies}).  This 
accounts for the small variation in occupations of these levels which 
is still evident in Fig.~3.

Fig.~4 compares with the experimental data of ref.~\cite{MITgrowth}, 
for two different temperatures.  The MIT group fitted their data to an 
uncorrected growth equation $\dot n_0 = \gamma n_0 [1 - (n_0/n_0({\rm 
eq}))^{2/5} ]$, and reported only values of the parameter $\gamma$.  
To represent that the experimental curve is a fitted curve rather than 
the raw data, the curve has been plotted as a broad band.  The MIT 
group used the initial population for their curves as free parameters 
in their fit.  We have set the initial populations for the MIT curves 
so as to give the best agreement with the initiation time of our 
growth.

\Beginfigure
\infig{8.5cm}{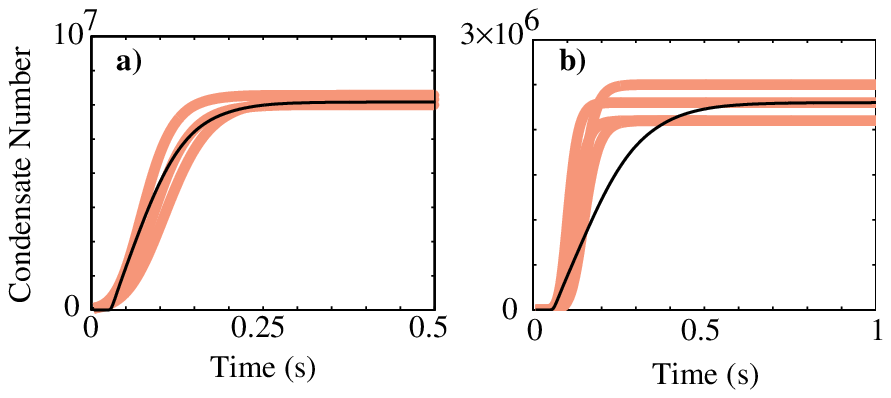}
\Caption{Fig.~4:  Comparison of theoretical growth curves (black) with
experimentally fitted curves (gray) from~\protect\cite{MITgrowth}.
{\bf a)}~Theory $T=830{\rm nK}$, $n_{0}({\rm eq})=7.6\times10^{6}$; 
Experiment  $T=810-890{\rm nK}$, $n_{0}({\rm eq})=7.5-7.85\times10^{6}$;
{\bf b)}~Theory $T=590{\rm nK}$, $n_{0}({\rm eq})=2.3\times10^{6}$; 
Experiment  $T=580-610{\rm nK}$, $n_{0}({\rm eq})=2.1-2.5\times10^{6}$.}
\Endfigure

In the $T=830{\rm nK}$ case, Fig.~4a, the growth speed predicted 
agrees with that experimentally found.  The $T=590{\rm nK}$ case in 
Fig.~4b shows a theoretical growth rate some three times slower than 
is measured.  The discrepancy between theory and experiment at lower 
temperatures is hard to evaluate using the data in the form presented 
in \cite{MITgrowth}, which do not allow for direct comparison between 
the actual projected spatial distributions as given by phase-contrast 
microscopy, and theoretical spatial distributions.  These are not 
difficult to calculate from our many-level growth curves---the 
methodology will be published elsewhere.  The MIT method fits to a 
zero chemical potential vapor plus a non-zero chemical potential 
condensate - a reasonable estimate in the absence of any theoretical 
description of the spatial distribution of the vapor.  But a detailed 
description might give quite different results for temperature, and 
for condensate and vapor numbers.


In summary, we have given a description of condensate growth which 
covers the full range of behaviors, both before and after initiation 
of the condensate.  It is {\em quantitative} and agrees quite well 
with experiment.  The behavior before initiation is essentially as 
given by the solution of the quantum Boltzmann equation, and agrees 
with computations based on this equation 
\cite{SVISTUNOV,Kagan92,Kagan97}.  But we are able to give a value for 
the initiation time which appears to be consistent with experiment.  
After the initiation of the condensate, the occupations of the 
non-condensate levels are clamped by their fast relaxation time to 
quasi-equilibrium values, which change with the rise in trap levels 
induced by the very much slower growth of the condensate to its final 
occupation.

 We believe the future 
development of the theory for this problem will involve mainly 
refinements of our picture, such as including depletion of our fixed 
``bath'' of vapor.  But for quantitative comparison with experiment, 
it will be necessary to have more extensive data on spatial 
distributions at a range of temperatures.

\noindent{\bf Acknowledgments:} We would like to thank Wolfgang  Ketterle and
Hans-Joachim Miesner for discussions on sodium experiments; Eric 
Cornell, Carl Wieman, Deborah Jin and Jason Ensher for discussions 
regarding rubidium experiments; and Yuri Kagan and Boris Svistunov for 
discussions regarding their kinetic approach.  This work was supported 
by the Marsden Fund under contract number PVT-603, and by the 
{\"O}sterreichische Fonds zur F{\"o}rderung der wissenschaftlichen 
Forschung.

This paper was developed from an earlier unpublished version 
\cite{NewerBosGro} during the program BEC-98, at the Institute for 
Theoretical Physics at Santa Barbara (supported by the National 
Science Foundation under grant number PHY94-07194). We
thank all involved in this program for the provision of a most 
stimulating environment.

\end{multicols}
\end{document}